\documentclass[referee]{aa} 
\usepackage{astron}
\usepackage{psfig}
\usepackage{graphics}

\psfigurepath{./PS}

\newcommand{\be}{\begin{eqnarray}}
\newcommand{\ee}{\end{eqnarray}}
\newcommand{\goto}{\rightarrow}
\newcommand{\bitem}{\begin{itemize}}
\newcommand{\eitem}{\end{itemize}}
\newcommand{\bR}{{\bf R}}

\newcommand{\cR}{{\cal R}}

\begin{document}
\title{Detection and discrimination of cosmological non-Gaussian signatures   
by multi-scale methods}
       
\author{J.-L. Starck\inst{1} \and N. Aghanim\inst{2} \and O. Forni\inst{2}}

\institute{ DAPNIA/SEDI-SAP, Service d'Astrophysique, CEA-Saclay, 
 F-91191 Gif-sur-Yvette Cedex, France
\and  IAS-CNRS, Universit\'e Paris Sud, B\^atiment 121, F-91405, Orsay Cedex, France }
 
\offprints{jstarck@cea.fr}
 
\date{\today}
% \date{{\bf to appear in Astronomy and Astrophysics}}

% \pagestyle{empty}
 
%============================================================================

\abstract{ Recent Cosmic Microwave Background (CMB) observations
indicate that the temperature anisotropies arise from
quantum fluctuations in the inflationary scenario. In the simplest
inflationary models, the distribution of CMB temperature fluctuations should be
Gaussian. However, non-Gaussian signatures can be present.
They might have different origins and thus different statistical
and morphological characteristics.

In this   context and motivated by   recent and future CMB 
experiments, we 
search for, and discriminate between, different 
non-Gaussian signatures. We analyse simulated maps of three
cosmological sources of temperature anisotropies: Gaussian distributed 
CMB anisotropies from inflation, temperature fluctuations from cosmic
strings and anisotropies due to the kinetic Sunyaev-Zel'dovich (SZ) effect
both showing a non-Gaussian character.
We use different multi-scale methods, namely, wavelet, ridgelet and 
curvelet transforms. 
%Ridgelets and curvelets take the form of basis elements which 
%exhibit very high directional sensitivity and are highly anisotropic, 
%which is not the case for wavelets. 
The sensitivity and the discriminating power of the methods is evaluated 
using simulated data sets. 

We find that the bi-orthogonal wavelet transform is the most powerful
for the detection of non-Gaussian signatures and that the curvelet and
ridgelet transforms characterise quite precisely and exclusively the
cosmic strings. They allow us thus to detect them in a mixture of CMB
+ SZ + cosmic strings.  We   show that not one method only should
be applied to understand non-Gaussianity but rather a set of different
robust and complementary methods should be used.  }

\maketitle 
\markboth{Detection
and Discriminating the Cosmological non-Gaussian Signatures}{}
%\titlerunning{Non-Gaussian Signatures Detection}
%\authorrunning{Starck et al.}

\keywords{Cosmology:cosmic microwave background:early universe,Methods: Data Analysis}

\section{Introduction}
The search for non-Gaussian signatures in the Cosmic Microwave
Background (CMB) temperature fluctuation maps  furnished by
MAP\footnote{http://map.gsfc.nasa.gov/} \cite{komatsu2003}, and to
be furnished by
PLANCK\footnote{http://astro.estec.esa.nl/SA-general/Projects/Planck/},
is of great interest for cosmologists.  Indeed, the non-Gaussian
signatures in the CMB can be related to very fundamental questions
such as the global topology of the universe \cite{riazuelo2002}, super
string theory, topological defects such as cosmic strings
\cite{bouchet88}, and multi-field inflation \cite{bernardeau2002}.  The
non-Gaussian signatures can, however, have a different but still
cosmological origin. They can be associated with the
Sunyaev-Zel'dovich (SZ) effect \cite{sunyaev80} (inverse Compton
effect) of the hot and ionised intra-cluster gas of galaxy clusters
\cite{gauss:aghanim99,cooray2001}, with the gravitational lensing by
large scale structures \cite{bernardeau2003}, or with the reionisation
of the universe \cite{gauss:aghanim99,castro2002}.  They may also be
simply due to foreground emission \cite{gauss:jewell01}, or to  
non-Gaussian instrumental noise and systematics \cite{banday2000}.

It has now become clear that the detection of
non-Gaussian signature is important and feasible (e.g.
\cite{detroia2003,komatsu2003}). Nevertheless, even if a non-Gaussian
signal is detected the question of its origin might still be posed
(e.g. \cite{banday2000}). Hence, {\it it remains to
separate between several non-Gaussian contributions.  This is
  the most crucial issue for the use  the
non-Gaussian signatures as cosmological tools. }

It is therefore not surprising that a large number of studies have
recently been devoted to the subject of the detection of non-Gaussian
signatures. Many approaches have been investigated such as the
Minkowski functionals and the morphological statistics
\cite{gauss:novikov00,gauss:shandarin02}, the bispectrum
\cite{gauss:bromley99,gauss:verde00,gauss:phillips01}, the trispectrum
\cite{gauss:kunz01}, or wavelet transforms
\cite{gauss:aghanim99,gauss:forni99,gauss:hobson99,gauss:barreiro01,gauss:cayon01,gauss:jewell01}. Different
wavelet methods have been studied, such as the \`a trous algorithm
\cite{starck:book98} and the bi-orthogonal wavelet transform
\cite{ima:mallat98}.

% Other multi-scale transforms, such the ridgelet transform or the
% curvelet transform, have been developed recently.
A series of recent papers
\cite{cur:candes99_1,Curvelets-StMalo,starck:sta01_3}, however, argued that 
wavelets and related
classical multi-resolution techniques are based on a limited dictionary
made up of roughly isotropic elements present at all scales and
locations. We view as a limitation the fact that those dictionaries
do not exhibit highly anisotropic elements and that there is only a
fixed number of directional elements, independent of scale.  
Despite the success of the classical wavelet viewpoint, there are objects,
such as filamentary structures  or cosmic strings, that do
not exhibit isotropic scaling and thus call for other kinds of
multi-scale representations. Following this, 
new multi-scale systems like curvelets and ridgelets \cite{Curvelets-StMalo}
have been introduced
which are very different from wavelet-like systems. Curvelets and
ridgelets take the  form of basis elements which exhibit very high
directional sensitivity and are highly anisotropic.  In
two dimensions, for instance, curvelets are localised along curves, in
three dimensions along sheets, etc.   
\par\bigskip
The goal of the present study is to compare these new multi-scale 
representations to standard wavelet methods  on a set of simulated 
maps representing  two generic families of non-Gaussianities 
(spherical-like and string-like sources), and to answer the following 
questions:
\begin{enumerate}
\item Are there multi-scale methods that are better suited to detect the 
non-Gaussian signatures in CMB data?
\item Can we  go beyond the detection and extract some information 
about the nature of the non-Gaussian signal?
\end{enumerate}

The second point is particularly important and it will represent
the heart of our present study. A
non-Gaussian character can be due to many effects: Topological
defects, the SZ effect, calibration problems, or even a combination of
them. Here, we perform a study based on the idea that a
given multi-scale transform is optimal to detect features which have
the shape of its basis elements (this is quite similar to the matched
filter approach).  Consequently, if different multi-scale methods are
able to detect a non-Gaussian signal, one of them will outperform the
others since the shape of the non-Gaussian features contained in the
data will be close to its basis elements. In particular, using
wavelet-like systems and curvelet and ridgelet-like systems, we expect
to differentiate between the spherical-like and string-like sources of
non-Gaussianity.

Sections~\ref{sect_transforms} and 3 briefly review the different
transforms and simulated maps considered in our analysis. We
present the analysis together with our main results in section 4 and
discuss them in section 5. We provide the main conclusions in
section 6.

\section{Multi-scale transforms}
\label{sect_transforms}

\subsection{Bi-orthogonal wavelet transforms}
The most commonly used wavelet transform algorithm is the
decimated bi-orthogonal wavelet transform (OWT). Using the OWT, a
signal $s$ can be decomposed as follows:
\begin{eqnarray}
 s(l) = \sum_{k} c_{J,k} \phi_{J,l}(k) 
       +  \sum_{k} \sum_{j=1}^J \psi_{j,l}(k) w_{j,k}
\end{eqnarray}
with $\phi_{j,l}(x) = 2^{-j} \phi(2^{-j}x-l)$ and $\psi_{j,l}(x) =
2^{-j} \psi(2^{-j}x-l)$, where $\phi$ and $\psi$ are respectively the
scaling and the wavelet functions.  $J$ is the number of resolutions
used in the decomposition, $w_{j}$ the wavelet coefficients (or
details) at scale $j$, and $c_{J}$ is a coarse or smooth version of
the original signal $s$. 
The present indexing is such that $j = 1$ corresponds to the
finest scale (high frequencies).

The two-dimensional algorithm is based on separate variables leading to
prioritising of horizontal, vertical and diagonal
 directions. The scaling function is defined by
$\phi(x,y) = \phi(x)\phi(y)$,  
and the detail signal is obtained from three wavelets:
\begin{itemize}
\item vertical wavelet :
$\psi^1(x,y) = \phi(x)\psi(y) $
\item horizontal wavelet:
$\psi^2(x,y) = \psi(x)\phi(y) $
\item diagonal wavelet:
$\psi^3(x,y) = \psi(x)\psi(y) $
\end{itemize}
which leads to three wavelet sub-images at each resolution level.
A given wavelet band is therefore defined by its 
resolution level $j$ ($j=1\dots J$) and its direction number $d$ 
($d=1\dots 3$, corresponding respectively to the horizontal, vertical, and
diagonal band).
  
% Thus, the algorithm outputs $3J+1$ sub-band
% arrays.  The present indexing is such that $j = 1$ corresponds to the
% finest scale (high frequencies).  Coefficients $c_{j,k}$ and $w_{j,k}$
% are obtained by means of the filters $h$ and $g$:
% \begin{eqnarray}
% c_{j+1,l} & = & \sum_k h(k-2l) c_{j,k} \nonumber \\
% w_{j+1,l} & = & \sum_k g(k-2l) c_{j,k}
% \end{eqnarray}
% where $h$ and $g$ verify:
% \begin{eqnarray}
% {1 \over 2} \phi({x \over 2}) & = &  \sum_k h(k) \phi(x-k) \nonumber \\
% {1 \over 2} \psi({x \over 2}) & = & \sum_k g(k) \phi(x-k)
% \end{eqnarray}
% and the reconstruction of the signal is performed with:
% \begin{eqnarray}
% c_{j,l} = 2 \sum_k [ \tilde h(k+2l) c_{j+1,k}  + \tilde g(k+2l) w_{j+1,k}  ]
% \end{eqnarray}
% where the filters $\tilde h$ and $\tilde g$ must verify the conditions of
% dealiasing and exact reconstruction:
% \begin{eqnarray}
% \hat{h}(\nu+{1\over 2}) \hat{\tilde h}(\nu) + \hat{g}(\nu+{1\over 2}) \hat{\tilde g}(\nu)  & = & 0 \nonumber \\
% \hat{h}(\nu) \hat{\tilde h} + \hat{g}(\nu) \hat{\tilde g}(\nu)  & = & 1.  
% \end{eqnarray}

% The application of the OWT to image compression, using
% the 7-9 filters \cite{wave:antonini92} and the 
% zerotree coding \cite{compress:shapiro93,compress:said96} 
% has lead to impressive results, compared to previous methods like JPEG.

% The undecimated version (UWT) of the OWT is certainly the most 
% popular transform for data filtering.
The OWT presents only a fixed number of directional elements 
independent of scales,
and there are no highly anisotropic elements \cite{cur:candes99_1}. 
For instance, the Haar 2D 
wavelet transform is optimal to find features with a  
ratio length/width $= 2$, and a horizontal, vertical, or diagonal 
orientation. Therefore, we naively expect the OWT to be optimal for 
detecting mildly isotropic or anisotropic features.
 
\subsection{The isotropic \`a trous wavelet transform}
The \`a trous wavelet transform algorithm 
decomposes an $n \times n$ image $I$ as a superposition of
the form
\[
I(x,y) = c_{J}(x,y) + \sum_{j=1}^{J} w_j(x,y),
\]
where $c_{J}$ is a coarse or smooth version of the original image $I$
and $w_j$ represents `the details of $I$' at scale $2^{-j}$ (see
Starck et al.\cite*{starck:book98,starck:book02} for more information). Thus, the
algorithm outputs $J+1$ sub-band arrays of size $n \times n$. (The
present indexing is such that $j = 1$ corresponds to the finest scale
(high frequencies)).

Hence, we have a {\em multi-scale pixel representation}, i.e. each
pixel of the input image is associated with a set of pixels of the
multi-scale transform. This wavelet transform is very well adapted to
the detection of isotropic features, and this explains  
its success for astronomical image processing, where the data contain
mostly isotropic or quasi-isotropic objects, such as stars, galaxies or
galaxy clusters.

\subsection{The ridgelet transform}
 The two-dimensional continuous ridgelet transform in $\bR^2$ can be
defined as follows \cite{cur:candes99_1}. We pick a smooth univariate
function $\psi:\bR \goto \bR$ with sufficient decay and satisfying the
admissibility condition
\begin{equation}
\label{eq:admissibility}
\int |\hat{\psi}(\xi)|^2/|\xi|^2 \, d\xi < \infty,
\end{equation}
which holds if, say, $\psi$ has a vanishing mean $\int
\psi(t) dt = 0$.  We will suppose that $\psi$ is normalised so that
$\int |\hat{\psi}(\xi)|^2 \xi^{-2} d\xi = 1$.

For each $a > 0$, each $b \in \bR$ and each $\theta \in [0,2\pi]$, we
define the bivariate {\em ridgelet} $\psi_{a,b,\theta}: \bR^2 \goto
\bR$ by
\begin{equation}
\label{eq:ridgelet}
       \psi_{a,b,\theta} ({\mathbf x}) = a^{-1/2} \cdot
     \psi( (x_1 \cos\theta + x_2 \sin\theta  - b)/a).
\end{equation}

Given an integrable bivariate function $f({\mathbf x})$, we define its
ridgelet coefficients by:
\[
\cR_f(a,b,\theta) = \int \overline{\psi}_{a,b,\theta}({\mathbf x}) f({\mathbf x}) d{\mathbf x}.
\]
We have the exact reconstruction formula
\begin{equation}
\label{eq:CRT}
f({\mathbf x}) = \int_0^{2\pi} \int_{-\infty}^\infty \int_0^\infty
\cR_f(a,b,\theta) {\psi}_{a,b,\theta}({\mathbf x}) \frac{da}{a^3} db
\frac{d\theta}{4\pi}
\end{equation}
valid for functions which are both integrable and square
integrable.

It has been shown \cite{cur:candes99_1} that the ridgelet transform is
precisely the application of a 1-dimensional wavelet transform to the
slices of the Radon transform.

\subsection*{Local ridgelet transform}
The ridgelet transform is optimal to find only lines of the size of
the image. To detect line segments, a partitioning must be introduced.
The image is decomposed into smoothly overlapping blocks of
side-length $B$ pixels in such a way that the overlap between two
vertically adjacent blocks is a rectangular array of size $B \times
B/2$; we use an overlap to avoid blocking artifacts. For an $n \times n$
image, we count $2n/B$ such blocks in each direction.  The
partitioning introduces redundancy, as a pixel belongs to 4
neighboring blocks.

More details on the implementation of the digital ridgelet transform
can be found in Starck et al\cite*{starck:sta01_3,starck:sta02_3}.  The ridgelet
transform is therefore optimal to detect lines of a given size, which
is the block size.

\subsection{The curvelet transform.}
\label{sec:curvelet}

The curvelet transform \cite{cur:donoho99,starck:sta02_3} opens   the
possibility to analyse an image with different block sizes, but with a
single transform. The idea is to first decompose the image into a set
of wavelet bands, and to analyse each band with a local ridgelet
transform. The block size can be changed at each scale level.  Roughly
speaking, different levels of the multi-scale ridgelet pyramid are
used to represent different sub-bands of a filter bank output.
 
The side-length of the localising windows is doubled {\em at every
 other} dyadic sub-band, hence maintaining the fundamental property of
 the curvelet transform, that elements of length about
 $2^{-j/2}$ serve for the analysis and synthesis of the $j$-th
 sub-band $[2^j, 2^{j+1}]$.  Note also that the coarse description of
 the image $c_J$ is not processed. In our implementation, we used the
 default block size value $B_{min} = 16$ pixels.  This implementation
 of the curvelet transform is also redundant. The redundancy factor is
 equal to $16J+1$ whenever $J$ scales are employed.  A given curvelet
 band is therefore defined by the resolution level $j$ ($j=1\dots J$)
 related to the wavelet transform, and by the ridgelet scale $r$.

This method is optimal to detect anisotropic structures of different
lengths.

\subsection{Statistics from the multi-scale coefficients}
Many kinds of statistics can be derived from the multi-scale
coefficients \cite{gauss:aghanim99}. In the following we use the
kurtosis of the coefficients obtained by the previously described
multi-scale transforms.  Several aspects have however to be considered.
 
\subsubsection*{The border problem}
Wavelet coefficients have been obtained by convolution with filters,
and coefficients close to the border of the image have therefore been
calculated using an extrapolation and should not be used when calculating
the excess kurtosis. For the bi-orthogonal wavelet transform, we use the 7/9
filter, which implies that coefficients at a distance from the border
smaller than three must not be taken into account.  For \`a trous
wavelet coefficients at scale $j$, the distance is $2^j$ (because it
is an undecimated wavelet transform).  For the same reasons, border
blocks in the ridgelet and the curvelet transforms must also not be
used.

\subsubsection*{Renormalisation of the ridgelet and the curvelet coefficients}
When using the ridgelet and the curvelet transforms, another problem
occurs.  Indeed, as a coefficient is obtained by integrating along a
given direction, the expectation value of a coefficient depends on
both the direction and the position of the line. For example,
coefficients relative to diagonal directions integrate more values
than those relative to horizontal and vertical directions. A
renormalisation is needed.  After applying the ridgelet transform
independently to all blocks, we obtained a set of $N_t$ transform
blocks $T_i(a,b,\theta)$ ($i=1 \dots N_t$), and for each scale,
orientation and position $(a,b,\theta)$, we extract the vector
$V_{a,b,\theta}(i)$.  For a stationary signal, dividing
$T_i(a,b,\theta)$ by the standard deviation of $V_{a,b,\theta}$ leads
to a good renormalisation, but it is not our case, (cosmological
signals are not stationary) and we therefore use a robust estimator,
$\mathrm{MAD}$ (Median Absolute Deviation), defined by
$\mathrm{MAD}(x) = \mathrm{median}(\mid x \mid )/0.6745$ \cite{astro:mad93}.
 Hence, we
normalise the ridgelet coefficients by the following expression:
\begin{eqnarray}
\bar{T}_i(a,b,\theta) = \frac{T_i(a,b,\theta)}{\mathrm{MAD}(V_{a,b,\theta})}.
\end{eqnarray}
For the curvelet transform, this renormalisation is performed
independently on each scale.

\section{Simulated astrophysical signals}
The temperature anisotropies of the CMB contain the contributions of
both the primary cosmological signal, directly related to the initial
density perturbations, and the secondary anisotropies.  The latter are
generated after matter-radiation decoupling \cite{white2002}.  They
arise from the interaction of the CMB photons with the neutral or
ionised matter along their path
\cite{sunyaev80,ostriker86,vishniac87}.

\begin{figure}[htb]
\centerline{
\hbox{
\psfig{figure=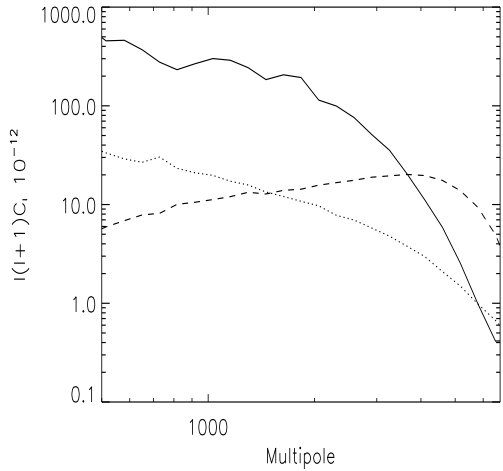,height=7cm,width=7cm}
}}
\caption{Power spectra of the different contributions used in this study. 
The solid line
represents the primary CMB anisotropies from an inflationary model 
(the fluctuations are Gaussian distributed). The dashed line is for the 
kinetic SZ effect from galaxy clusters. The dotted line stands for the 
power spectrum of the allowed contribution from cosmic strings, i.e.
with the parameter $\alpha=0.82$.
}
\label{fig:spec}
\end{figure}

In the present study, we assume that the primary CMB anisotropies are
dominated by the fluctuations generated in the simple single field
inflationary Cold Dark Matter model with a non-zero cosmological
constant. The CMB anisotropies have therefore a Gaussian distribution. We
allow for a contribution to the primary signal from topological
defects, namely cosmic strings (CS), as suggested in
\cite{gauss:bouchet00}. In addition, we take into account the
secondary anisotropies due to the kinetic Sunyaev-Zel'dovich (SZ)
effect \cite{sunyaev80}.  The SZ effect represents the Compton
scattering of the CMB photons by the free electrons of the ionised and
hot intra-cluster gas. When the galaxy cluster moves with respect to
the CMB rest frame, the Doppler shift induces additional anisotropies;
this is the so-called kinetic SZ (KSZ) effect. As the latter
fluctuations have the same spectral signature as the primary, we add
the two signals directly. The kinetic SZ maps are simulated following
Aghanim et al \cite*{gauss:aghanim01b}. We use for our simulations the cosmological
parameters obtained from the WMAP satellite \cite{astro:bennett2003}
and a normalisation parameter $\sigma_8=0.9$. Finally, we obtain 
the so-called ``simulated observed map'', $D$, that contains the three
previous astrophysical components. It is obtained from
$D=\sqrt{\alpha}\mathrm{CMB}+\sqrt{1-\alpha}\mathrm{CS}+\mathrm{KSZ}$,
where $\mathrm{CMB}$, $\mathrm{KSZ}$ and $\mathrm{CS}$ are
respectively the CMB, the kinetic SZ and the cosmic string simulated
maps.  $\alpha=0.82$ is a constant derived by
\cite{gauss:bouchet00}. All the simulated maps  
have $500 \times 500$ pixels with a resolution of
 1.5 arcminute per pixel.  Apart
from the dominant inflationary component which is Gaussian, all the
other contributions are non-Gaussian. However, KSZ and
CS have different characteristics as the KSZ and CS induce
respectively spherical-like and string-like structures in the CMB. In
order to illustrate how difficult  the task of detecting and
separating different non-Gaussian signatures is, we show in
Fig.~\ref{fig_cmb} a set of simulated maps. Primary CMB, kinetic SZ
and cosmic string maps are shown respectively in Fig. \ref{fig_cmb}
top left, top right and bottom left. The ``simulated observed
map", containing the three previous components, is displayed in
Fig. \ref{fig_cmb} bottom right. It can easily be seen from Fig.
\ref{fig:spec}, which displays the power spectra of the different
components, that the primary CMB anisotropies dominate all the
signals except at very high multipoles (very small angular scales).

\begin{figure*}[htb]
\vspace{6cm}
% \centerline{
% \hbox{
% \psfig{figure=fig_cmb.ps,bbllx=1.9cm,bblly=12.7cm,bburx=14.6cm,bbury=25.4cm,height=7cm,width=7cm,clip=}
% \psfig{figure=fig_sz.ps,bbllx=1.9cm,bblly=12.7cm,bburx=14.6cm,bbury=25.4cm,height=7cm,width=7cm,clip=}
% }}
% \centerline{
% \hbox{
% \psfig{figure=fig_cs.ps,bbllx=1.9cm,bblly=12.7cm,bburx=14.6cm,bbury=25.4cm,height=7cm,width=7cm,clip=}
% \psfig{figure=fig_cmb_sz_cs.ps,bbllx=1.9cm,bblly=12.7cm,bburx=14.6cm,bbury=25.4cm,height=7cm,width=7cm,clip=}
% }}
\caption{Top, primary Cosmic Microwave Background anisotropies (left) and 
kinetic Sunyaev-Zel'dovich fluctuations (right).
Bottom, cosmic string simulated map (left) and simulated observation
containing the previous three components (right).}
\label{fig_cmb}
\end{figure*}

\section{Analysis and results}
\subsection{Sky component multi-scale analysis.}
On each of the three sets of maps (i.e. CMB, CS and KSZ), we have 
first run the following multi-scale methods: 
\begin{itemize}
\item \`a trous wavelet transform.
\item bi-orthogonal wavelet transform, using the standard 7/9 filter 
\cite{wave:antonini92}.
\item local ridgelet transform with a block size equal to 16 pixels.
\item local ridgelet transform with a block size equal to 32 pixels.
\item curvelet transform with a block size equal to 16 pixels.
\end{itemize}
This processing has been repeated on 100 CMB and 100 KSZ simulated
maps and on the three available CS maps.  In order to estimate the
sensitivity of each transform to the different types of non-Gaussian
signatures associated with the different components individually
(sharp edges for CS and point sources for KSZ), we have calculated the
mean excess kurtosis (fourth moment) and its standard deviation on
each individual resolution level.  In this analysis, we do not present
the skewness (third moment) since we do not expect the KSZ signal to
be ``skewed''\cite{astro:dasilva2002}. The comparison of the
transforms on the basis of this non-Gaussian estimator in our case
would be meaningless.

For example, the \`a trous wavelet transform applied to the CMB
produces 100 individual values for the excess of kurtosis per
scale (resolution level). We note $K_a(i,b)$ the excess kurtosis of
the $i$th simulated CMB map ($i=1 \dots 100$) at the band $b$.
For the \`a trous wavelet transform and the ridgelet transform, 
we have $b=j$, where $j$ is the scale ($j = 1 \dots J$) 
and $J$ is number of scales of the wavelet transform (we
used $J = 5$ in our experiments). For a bi-orthogonal wavelet
transform, we have three bands per scale ($b = 1 \dots 3J$),
and $b$ can also be represented by the two indices $(j,d)$, 
$d$ ($l = 1 \dots 3$) standing respectively for the horizontal, vertical or
diagonal directions. The same holds for the curvelet transform where $b$ can 
 be represented by the two indices $(j,r)$ where $r$ is the scale index 
in the ridgelet transform. 

We derive from the ensemble of $K_a$ obtained on the CMB maps the mean
kurtosis value ${\bar K_a}(b)$ and the standard deviation
$\sigma_{K}(b)$. Table~\ref{tab_kur_cmb} gives the maximum value of
${\bar K_a}(b)$ (among all bands), its associated scale and the
standard deviation (columns 1, 2, and 3). It also gives the maximum
value for the ratio ${\bar K_a}(b)/\sigma_{K_G}$, where $\sigma_{K_G}$
is the standard deviation of the kurtosis in the Gaussian set of maps,
as well as the associated scale (columns 4 and 5).  The same treatment
is applied on the KSZ and the CS components.  The results are
respectively given in Tables~\ref{tab_kur_sz} and \ref{tab_kur_cs}.

{\small
\begin{table}[htb]
\baselineskip=0.4cm
\begin{center}
\begin{tabular}{lcccc|ccc} \hline \hline
Multi-scale Method              & ${\bar K_a}$  & $\sigma_K$ & Scale&  & $\frac{\bar K_a}{\sigma_{K_G}}$  &   Scale \\ \hline \hline
Bi-orthogonal Wavelet           & 0.006  &  0.098 & 3,2&  & 0.262 & 1,3 \\
\`a trous WT                    & -0.04  &  0.198 & 5 & & 0.20 & 3\\
Local ridgelets ($B = 16$)      & 0.009  &  0.028 & 1 & & 0.33 & 1\\
Local ridgelets ($B = 32$)      & 0.061  &  0.030 & 1 & & 1.976 & 1 \\
Curvelets ($B = 16$)            & 0.271  &  0.183 & 5,3 & & 3.540 & 4,1 \\ \hline
\hline
\end{tabular}
\caption{Table of the maximum value among all bands of the mean excess 
kurtosis ${\bar K_a}$ (column 1), its associated scale (column 3) and the 
standard deviation $\sigma_K$ (column 2) for the primary CMB anisotropies 
(inflationary model) only. Columns 4 and 5 give the maximum
value for ${\bar K_a}(b)/\sigma_{K_G}$ and the
associated scale}
% \vspace{0.5cm}
\label{tab_kur_cmb}
\end{center}
% \end{table}
% }
% {\small
% \begin{table*}[htb]
% \baselineskip=0.4cm
\begin{center}
\begin{tabular}{lcccc|ccc} \hline \hline
Multi-scale Method               & ${\bar K_a}$ & $\sigma_K$  & Scale &  & $\frac{\bar K_a}{\sigma_{K_G}}$  & Scale \\ \hline \hline
Bi-orthogonal Wavelet           & 46.673  &  35.26  &  1,1 & &  2081.41 &  1,3 \\
\`a trous WT                    & 80.380  &  55.898 &  1  & & 4354.48 &  1 \\
Local ridgelets ($B = 16$)      & 18.961  &  8.886 &   2  & & 397.85  &  2 \\
Local ridgelets ($B = 32$)      & 14.715  &  8.446 &  3  &  & 229.01  &  2 \\
Curvelets ($B = 16$)            & 12.360  &  9.475 &  3,4 &  & 351.80  & 1,1 \\ \hline
\hline
\end{tabular}
\caption{Table of the maximum value among all bands of the mean excess 
kurtosis ${\bar K_a}$ (column 1), its associated scale (column 3) and the 
standard deviation $\sigma_K$ (column 2) for the kinetic SZ fluctuations only.
Columns 4 and 5 give the maximum
value for ${\bar K_a}(b)/\sigma_{K_G}$ and the
associated scale.}
% \vspace{0.5cm}
\label{tab_kur_sz}
\end{center}
% \end{table*}
% }
% {\small
% \begin{table*}[htb]
% \baselineskip=0.4cm
\begin{center}
\begin{tabular}{lcccc|ccc} \hline \hline
Multi-scale Method              & ${\bar K_a}$ & $\sigma_K$ & Scale & & $\frac{\bar K_a}{\sigma_{K_G}}$  & Scale  \\ \hline \hline
Bi-orthogonal Wavelet           & 51.36  &  1.90 & 1,2 & & 2059.74 & 1,2\\
\`a trous WT                    & 35.76  &  7.99 & 1    &  & 1937.78 & 1 \\
Local ridgelets ($B = 16$)      & 6.98   &  0.95 & 1     & & 243.02 & 1   \\
Local ridgelets ($B = 32$)      & 3.69   &  0.46 & 1     & & 117.67 & 1 \\
Curvelets ($B = 16$)            & 10.43  &  1.40 & 1,1   & &  609.06 &  1,1 \\ \hline
\hline
\end{tabular}
\caption{Table of the maximum value among all bands of the mean excess 
kurtosis ${\bar K_a}$ (column 1), its associated scale (column 3) and the 
standard deviation $\sigma_K$ (column 2) for the cosmic strings only. 
Columns 4 and 5 give the maximum
value for ${\bar K_a}(b)/\sigma_{K_G}$ and the
associated scale. Note that only 3 maps are analysed in this case.}
% \vspace{0.5cm}
\label{tab_kur_cs}
\end{center}
\end{table}
}
 
As it was expected, the \`a trous algorithm is better suited to detect
isotropic features such as the ``quasi-spherical'' KSZ anisotropies
due to galaxy clusters. Indeed, we note from Table ~\ref{tab_kur_sz}
that the multi-scale excess kurtosis for KSZ is largest when using
the \`a trous algorithm. Unexpectedly though, the multi-scale excess
kurtosis for CS (Table \ref{tab_kur_cs}) is largest when using the
bi-orthogonal wavelet transform whereas the excess kurtosis from the
ridgelet and the curvelet transforms show that both transforms seem
not that well adapted to test the CS non-Gaussian signatures.  An
explanation for this behaviour is that the CS (see Fig.~\ref{fig_cmb})
  presents not only  elongated structures and sharp edges, but also a
large number of spots and spherical-like structures along the edges
and at the intersection of edges, which contribute significantly to
the non-Gaussian character. These spots are very well detected by the
wavelets.
 
% addition, the non-zero value for the excess kurtosis for the CMB coefficients
% (in particular for the ridgelet transform) may, at face value, lead to some 
% difficulties to interpret the results in terms of their statistical character.
% This behaviour (non-zero excess kurtosis) is not quite surprising. It was 
% indeed already noted by 
% \cite{gauss:aghanim99} for the OWT analysis. In fact, it is due to the rapid 
% variation of power in the CMB, namely to the cut off at small scales 
% of the power spectrum. The proper statistical characterisation of the signal
% is based on the comparison of the non-Gaussian maps of a given signal with 
% a set of Gaussian realisations with the same power
% spectrum. In this case, this behaviour does therefore not affect the
% interpretation as discussed in the previous studies.

\subsection{Testing the sensitivities to Gaussian+non-Gaussian signals}

In this section, we study the relative sensitivity of the different
multi-scale transforms to the two families of different non-Gaussian
characters when the signals are added to a dominant Gaussian noise,
i.e.  the primary CMB.  We use three datasets $D^{(1)}$,
$D^{(2)}$ and $D^{(3)}$.  We have created 300 simulated maps by adding
the 100 CMB realisations to the KSZ ($D^{(1)}_i = \mathrm{CMB}_i +
\mathrm{KSZ}$, and $i = 1 \dots 100$), to the CS ($D^{(2)}_i =
\sqrt{\alpha}\mathrm{CMB}_i + \sqrt{1 - \alpha}\mathrm{CS}$), and to
both the KSZ and CS components ($D^{(3)}_i =
\sqrt{\alpha}\mathrm{CMB}_i + \sqrt{1 - \alpha}\mathrm{CS} +
\mathrm{KSZ}$).  Then we apply our five multi-scale transforms to
these 300 maps.  As in the previous section, we have calculated for
each band $b$ of each transform and for each dataset $D^{(l)}$
($l=1,2,3$) the mean kurtosis value ${\bar K}_{D^{(l)}}(b)$ and the
standard deviation $\sigma_{K_{D^{(l)}}, b}$. In order to calibrate
and compare the departures from a Gaussian distribution, we have
simulated for each image $D^{(l)}_i$ a Gaussian Random Field
$G^{(l)}_i$ which has the same power spectrum as $D^{(l)}_i$. This
allows us to calculate a normalised mean kurtosis given by:
\begin{eqnarray}
{\cal K}_{l}(j) = \frac{ {\bar K}_{D^{(l)}}(b) - {\bar K}_{G^{(l)}}(b)}
 { \sigma_{K_{G^{(l)}}, b} }
\end{eqnarray}

We give in Tables~\ref{tab_kur_cmb_sz},~\ref{tab_kur_cmb_cs}
and~\ref{tab_kur_cmb_sz_cs} the results in terms of the maximum values for
illustration. Complete Tables can be found in the Appendix 
% complete Tables be provided upon request 
for the three datasets $D^{(1)}$, $D^{(2)}$ and $D^{(3)}$. The first
three columns represent maximum ${\bar K}_{a}$, standard deviation and
associated scale. The last two columns give the maximum
normalised mean kurtosis and associated scale.

{\small
\begin{table}[htb]
\baselineskip=0.4cm
\begin{center}
\begin{tabular}{lcccc|ccc} \hline \hline
Multi-scale Method               & ${\bar K_a}$ &$\sigma_K$ & Scale  & & ${\cal K}$  & Scale  \\ 
\hline \hline
Bi-orthogonal Wavelet           & 22.94  &  0.098  & 1,3 & & 1106.58 & 1,3 \\
\`a trous WT                    & 0.73  &  0.06 & 1 & & 65.79 & 1 \\
Local ridgelets ($B = 16$)      & 0.013  &  0.029  & 1&  & 0.124 & 1  \\
Local ridgelets ($B = 32$)      & 0.062  &  0.030 & 1 & & 0.114 & 1 \\
Curvelets ($B = 16$)            & 0.276  &  0.109 &  4,3  & & 10.12 & 1,1  \\ \hline
\hline
\end{tabular}
\caption{Table of maximum mean excess kurtosis ${\bar K_a}$, and associated 
standard deviation $\sigma_K$  and scale (first three 
columns). The last two columns give the maximum normalised mean kurtosis 
${\cal K}$ and associated scale for $D^{(1)}=$CMB+KSZ. }
% \vspace{0.5cm}
\label{tab_kur_cmb_cs}
\end{center}
\begin{center}
\begin{tabular}{lcccc|ccc} \hline \hline
Multi-scale Method               & ${\bar K_a}$ &$\sigma_K$ & Scale & &  ${\cal K}$  & Scale   \\ 
\hline \hline
Bi-orthogonal Wavelet           &  37.96  & 0.14   &  1,3 & & 1813.61 & 1,3  \\
\`a trous WT                    &  5.74  &  0.15  & 1& & 424.15 &  1 \\
Local ridgelets ($B = 16$)      &  0.174  & 0.033  & 1& & 5.68 &  1 \\
Local ridgelets ($B = 32$)      &  0.152 &  0.031 &  1& & 2.84 & 1 \\
Curvelets ($B = 16$)            &  2.22  &  0.055  &  1,1& & 198.62 & 1,1  \\ \hline
\hline
\end{tabular}
\caption{Table of maximum mean excess kurtosis ${\bar K_a}$, and associated 
standard deviation $\sigma_K$  and scale (first three 
columns). The last two columns give the maximum normalised mean kurtosis 
${\cal K}$ and associated scale for $D^{(2)}=$CMB+CS. }
% \vspace{0.5cm}
\label{tab_kur_cmb_sz}
\end{center}

\begin{center}
\begin{tabular}{lcccc|ccc} \hline \hline
Multi-scale Method               & ${\bar K_a}$ &$\sigma_K$ & Scale &&  ${\cal K}$  & Scale  \\ 
\hline \hline
Bi-orthogonal Wavelet           & 26.22  &  0.53 & 1,2 && 1040.54& 1,2  \\
\`a trous WT                    & 5.20  &  0.12 & 1 && 392.25 &  1 \\
Local ridgelets ($B = 16$)      & 0.17  &  0.032 & 1 & & 5.88 & 1 \\
Local ridgelets ($B = 32$)      & 0.149  &  0.030 & 1 & & 2.99 & 1 \\
Curvelets ($B = 16$)            & 1.79  &  0.046 & 1,1 & & 165.68 & 1,1 \\ \hline
\hline
\end{tabular}
\caption{Table of maximum mean excess kurtosis ${\bar K_a}$, and associated 
standard deviation $\sigma_K$  and scale (first three 
columns). The last two columns give the maximum normalised mean kurtosis 
${\cal K}$ and associated scale for $D^{(3)}=$CMB+KSZ+CS. }
% \vspace{0.5cm}
\label{tab_kur_cmb_sz_cs}
\end{center}
\end{table}
}

From Tables~\ref{tab_kur_cmb_cs},~\ref{tab_kur_cmb_sz}
and~\ref{tab_kur_cmb_sz_cs}, we note that:
\begin{itemize}
\item Whatever the transform, it is at the first resolution level that
the non-Gaussian character is the best detected.  
\item The third band of the bi-orthogonal wavelet transform
(i.e. diagonal details of the first wavelet scale) is generally the
best for detecting non-Gaussian signatures in the CMB, even if the
structures are isotropic and spherical-like (see previous section).
\item The most important result is that  the 
normalised mean kurtosis of the ridgelets is compatible with 0 in the
case CMB+KSZ whereas it is non-zero in the case CMB+CS. Moreover, it 
conserves the same value in the case CMB+SZ+CS. For the normalised
mean kurtosis of the curvelets there is a non-zero value for the CMB+KSZ 
signal but it is almost 20 times smaller than in the cases CMB+CS and 
CMB+SZ+CS.
\end{itemize}
The first remark is related to the relative contributions of the
different astrophysical components. The primary CMB dominates
over the other components except at the first decomposition (or
wavelet) scale as can be seen from figure \ref{fig:spec}.  At this
scale corresponding typically to 3 arcminutes, the contribution from
the SZ effect and the CS is of the order of, or dominates, the primary
CMB. The non-Gaussian signatures introduced by both KSZ and CS are thus
easier to detect at this scale. It is worth noting however, that the
non-Gaussian character remains detectable at larger scales (second and
third decomposition scales).
 
The second point was already mentioned in Aghanim and Forni\cite*{gauss:aghanim99}.
In the previous section we have shown that the spherical-like
sources of non-Gaussianity were better detected by the \`a trous
transform when they are not mixed with a dominant Gaussian signal (the
CMB here). This property is not conserved when the quasi-spherical
non-Gaussianities such as the KSZ effect are added to the primary
CMB. The reason why the so-called diagonal details are more
sensitive than the other tests is the following: In the Fourier
domain, we note $R(u,v) = \frac{S(u,v)}{C(u,v)}$, the Signal-to-Noise
ratio (SNR), where $S$ is the power spectrum of the non-Gaussian
signal and $C$ is the power spectrum of the primary CMB (Gaussian
signal in our case). As the CMB power spectrum is isotropic, we have
$R(u,v) = \frac{S(u,v)}{C(\sqrt{u^2+v^2})}$.  When $\rho =
\sqrt{u^2+v^2}$ increases, the SNR increases, which explains why the
first wavelet scale presents the highest sensitivity to the
non-Gaussian features. Additionally, it becomes clear that $R(u,v)$ is
higher for $u=v=N/2$ rather than for $u=0$ and $v=N/2$ or $u=N/2$ and
$v=0$, which implies that the diagonal band (or details) is more
sensitive than horizontal and vertical bands, and confirms the results
described in Aghanim and Forni \cite*{gauss:aghanim99}.

The third point is undoubtedly the most important result of
this comparison. It shows that the normalised mean kurtosis of the
ridgelets is able to highlight the non-Gaussian character of the CS
buried in the CMB Gaussian signal with a mixing ratio of 0.18 (in
power).  {\it The normalised mean kurtosis of the
ridgelets are sensitive only to the string-like
non-Gaussianities}. The numbers are the same for the CMB+CS and
CMB+SZ+CS cases and there is no detection in the CMB+KSZ.  A similar
behaviour is also noticeable for the normalised mean kurtosis of the
curvelets. In this case, there is a detection of the
non-Gaussian character associated with the KSZ effect which is
one order of magnitude smaller than in the case of CMB+SZ+CS. These
behaviours discussed at the first decomposition scale are true for
larger scales ($j=2$).

From this last remark, we already see that we have found a family of
multi-resolution transforms, namely anisotropic based-systems
(ridgelets and curvelets) that are sensitive to a unique family of
non-Gaussian signatures, namely the string-like structures. In the
following section, we propose a strategy based on our results to
discriminate between spherical-like and string-like contributions to
the non-Gaussian signatures.

\subsection{Discriminating between the non-Gaussian signatures}
If a non-Gaussian signature is detected in the CMB by a given method,
it can be due to calibration problems or to several astrophysical
components such as KSZ or CS. Therefore, an excess kurtosis may be very
difficult to interpret in terms of its origin. While some limits can
be put on the excess due to instrumental or calibration problems, the
discrimination between astrophysical components is not obvious.

We investigate in this section a method by which the use of several
multi-scale transforms helps us in understanding the nature of the
detected non-Gaussian features.  In particular, we focus on the
possibility to discriminate the case 'CMB+SZ' from the case
'CMB+SZ+CS'. From tables~\ref{tab_kur_cmb_cs},~\ref{tab_kur_cmb_sz}
and~\ref{tab_kur_cmb_sz_cs}, we can easily see that the wavelet
transforms are sensitive to both the KSZ and the CS, while the
ridgelet transform and the curvelet transform are not or are less
sensitive to the KSZ. However, the numbers we obtain for the
normalised mean excess kurtosis are of the order of a few in the best
cases. It is thus important to find a better estimator of the
non-Gaussian signatures. We require an estimator that enhances the
signal to noise ratio. The product of the
normalised kurtosis obtained by two different transforms is a
promising approach for our problem. We note ${\cal K}_{AT}, {\cal
K}_{CUR}, {\cal K}_{R16}, {\cal K}_{R32}$ the normalised kurtosis
using the \`a trous wavelet transform, the curvelet transform, the
ridgelet transform with a block size equal to 16 and the ridgelet
transform with a block size equal to 32 respectively.
Table~\ref{tab_kur_mult} gives such products. The first column
indicates the two band numbers used, the last three columns
give the value of the product for the cases CMB+CS, CMB+SZ and
CMB+SZ+CS respectively.  We have given illustrative
values. The whole set of values can be found in the Appendix (Table
\ref{taball_kur_mult}).

A significant detection of non-Gaussianity by
the product of wavelet-like and curvelet-like transforms must rely on
comparable bands. More specifically, the curvelet bands noted $(j,r)$
for which $j$ and $r$ are too different, are meaningless by
construction.  Moreover, for the largest decomposition levels, the
number of coefficients are small and the results are thus more
sensitive to sample variance.

We see from these values that the product of the normalised kurtosis
of the wavelet transform (here the \`a trous transform for
illustration\footnote{Similar results are obtained for the OWT
transform but in this case there are 3 bands per scale and  many more
combinations and products are possible. The associated numbers can be
obtained upon request}) by normalised kurtosis of the curvelet
transform clearly discriminates between the case CMB+SZ and
CMB+SZ+CS. This is true up to the second decomposition scale. 

{\small
\begin{table}[htb]
\baselineskip=0.4cm
\begin{center}
\begin{tabular}{lcccc} \hline \hline
Kurtosis product                  & bands & CMB+CS & CMB+SZ  &  CMB+SZ+CS   \\ \hline \hline
${\cal K}_{AT} * {\cal K}_{CUR}$ & 1- 1,1  & 84248. & 665.8  &  64990.1   \\
                                 & 1- 1,2  & 2225.8 & 0.702  &  2163.28   \\
				 & 2- 2,2  & 57.74  & 0.013    &  66.024  \\
				 & 2- 2,3  & 7.97  & 0.093     &   8.58   \\
${\cal K}_{AT} * {\cal K}_{R16}$ & 1-1  &  2410.19  & 8.21  &  2307.2  \\
				 & 2-2  &     8.50  & 0.082  &  9.465    \\
${\cal K}_{AT} * {\cal K}_{R32}$ & 1-1  & 1204.6    &  3.978 &   1172.0 \\
                                 & 2-2  &  4.547     &  0.139 & 4.692\\
                                 & 3-3  &  0.017    &   0.001 & 0.004 \\ \hline
\hline
\end{tabular}
\caption{Product of the normalised excess kurtosis.}
% \vspace{0.5cm}
\label{tab_kur_mult}
\end{center}
\end{table}
}

\section{Discussion}

We tackle the problem of
the separation between different sources of non-Gaussian
signatures. We are not yet able   to identify the exact contribution of
the different effects but rather we detect that there are two families
of non-Gaussian features.

In ``real-life'' we will of course be a priori unable to separate
perfectly the two contributions, KSZ and CS, from the primary
CMB. Tables such as  Table \ref{tab_kur_mult} will therefore only
contain the numbers associated with the case CMB+CS+KSZ and no
relative detection will be made. However, the simulations  
allow us to calibrate the method.
We show (Table \ref{tab_kur_mult}) that there can be
orders of magnitudes of difference in terms of the products of normalised
kurtosis between cases where CS are present and cases where they are not.
We can therefore conclude from our study that whenever the product of
the normalised kurtosis from the wavelets and from the curvelets (or
ridgelets) are of the order of a few this indicates that string-like
non-Gaussian features are present in the total signal.  A larger
number would allow us to fully characterise the non-Gaussian
properties of the signals and to calibrate the method.

At this stage, the precise effects of the instrumental noise and of
the beam convolution are not taken into account and performing such
an analysis is beyond the scope of our study.  However,  
the non-Gaussian signals are detected not only at the
first decomposition scale but also at larger scales (up to $j=3$).
The beam convolution is likely to affect the first and second scale
but   present and future instruments (Planck, ACT, SPT, ACBAR) will
have better angular resolution and the beam effects are likely to be
less important.  As for the noise, if we assume it Gaussian (which
is the case in most studies) it will make the non-Gaussian signatures
more difficult to detect by reducing the overall signal to noise
ratio.  However, the numbers presented in Table \ref{tab_kur_mult} show
that the detection of CS is quite significant and we do not expect it
to vanish if white noise is added unless the noise has similar
features as a CS. More generally, in the case of any systematic effect
exhibiting anisotropic structures our method will detect them without
a priori being able to disentangle their origin from a cosmological
origin like the CS.
% Finally, both beam and noise effects should be studied and
% characterised. We believe that, provided the shape of the beam and the
% noise level, optimal bases might be found in this case.

In this   analysis, we present only the excess kurtosis
as an estimator of non-Gaussianity. We perform a
comparison of the multi-scale transforms and of their sensitivity to
the signals on the basis of the non-Gaussian character  to
highlight different contributors. Since we do not expect the KSZ
signal to be ``skewed'', a comparison based on the skewness (third
moment) would be meaningless. As expected, our simulations have shown
that the skewness is indeed not as sensitive as the kurtosis.

\section{Conclusion}
In the present study, we have tested the relative sensitivities of the
commonly used multi-scale transforms to the detection of  
non-Gaussian signatures in CMB maps. We have used simulated
maps of the primary anisotropies (assumed to be Gaussian) and we have taken
into account the contributions from the kinetic SZ effect and the
topological defects (cosmic strings); each of them represents one
family of non-Gaussian contributors (spherical-like and
string-like). The main results are:

\begin{itemize}
\item the bi-orthogonal wavelet transform is the  
most sensitive to the non-Gaussian signatures associated 
with cosmic strings buried in the primary Gaussian CMB signal.
% in terms of detection of the non-Gaussian signatures, we
% confirmed that the diagonal details of a bi-orthogonal wavelet 
% transform (OWT) are the most sensitive to non-Gaussianity.
% We were able to explain the reason for
% it which is related to the behaviour of the studied signals in the Fourier
% domain. We have shown that the OWT
% is the best tool to detect the non-Gaussian signatures associated 
% with cosmic strings buried in the primary Gaussian CMB signal.
\item The ridgelet and curvelet transforms are well adapted to {\it
discriminate} between string-like and quasi-spherical like non-Gaussian
features.
% Surprisingly although their anisotropic nature, we find that the
% ridgelet and curvelet transforms does not appear the most powerful
% tool to detect the non-Gaussian signatures associated with the cosmic
% strings. However, these transforms are quite adapted to {\it
% discriminate} between string-like and quasi-spherical like non-Gaussian
% features.
\end{itemize} 

In order to study the non-Gaussian signatures in the CMB and use them
as a cosmological tool to probe the early universe or the cosmic structures,
it is not sufficient to detect them accurately. The most important step is
to be able to separate the different contributions to the signal. In this
context, we clearly show that not only one method should be applied 
but rather a set of different robust and well understood methods.

In the present study, we  use several multi-scale
transforms: The isotropic wavelet transforms suited for spherical-like
sources of non-Gaussianity, and a curvelet transform  representing well
 sharp and elongated structures. Each  provides  
  an adapted non-Gaussian estimator, namely the normalised mean
excess kurtosis. We show that the combination of these transforms
through the product of the normalised mean excess kurtosis of isotropic
wavelet transforms by normalised mean excess kurtosis of curvelet
transforms highlights the presence of the cosmic strings in a mixture
CMB+KSZ+CS. Such a combination   gives information about the
nature of the non-Gaussian signals.   
Even though the detection power of the non-Gaussian character by the
ridgelets and curvelets is not important compared to that of
the wavelet, their sensitivity to a particular shape makes them a very
strong discriminating tool.

This is a first step towards the separation of the statistical
contributions to the CMB signal and it can help in the more general
context of component separation.

\begin{acknowledgements}
The cosmic string maps were kindly provided by F.R. Bouchet. We wish
to thank David Donoho for useful discussions. We are grateful to the
referee for helpful comments on an earlier version. 
\end{acknowledgements}

% \bibliographystyle{astron}
% \bibliography{gauss,candes,entropy,starck,wave,restore,ima,astro,compress,mc,curvelet,nab}

\section*{Annex A:}
{\tiny
\begin{table}[htb]
\baselineskip=0.4cm
\begin{center}
\begin{tabular}{lcccc} \hline \hline
Multi-scale Method      &   Scale  &  ${\bar K_a}$ &$\sigma_K$ & ${\cal K}$   \\ 
\hline \hline

Bi-orthogonal Wavelet   &  1,1   &   6.305    &   0.188 &   275.096  \\
     &  1,2   &   7.205    &   0.270 &     314.839  \\
     &  1,3   &  22.942    &   0.098 &      1106.586 \\
     &  2,1   &   0.068    &   0.058 &      1.788 \\
     &  2,2   &   0.057    &   0.056 &      1.416 \\
     &  2,3   &     1.782  &   0.151 &     47.186 \\
     &  3,1   &   0.013    &   0.088 &     0.134 \\
     &  3,2   &  0.009     &   0.098 &    0.040 \\
     &  3,3   & -0.002     &   0.082 &   0.090  \\
     &  4,1   &  -0.038    &   0.180 &    0.013 \\
     &  4,2   &  -0.014    &   0.184 &   0.008  \\
     &  4,3   & -0.005     &   0.175 &    0.064 \\
     &  5,1   &  -0.088    &   0.323 &    0.011 \\
     &  5,2   &   -0.111   &   0.466 &   0.023  \\
     &  5,3   &  -0.042    &   0.530 &  0.023   \\
\`a trous WT   &    1    &     0.731  &      0.060  &  65.793  \\
       &   2   &  0.018  &     0.031  &      1.209  \\
       &   3   &  -0.006 &     0.048  &    0.029  \\
       &   4   &  -0.017 &     0.101  &    0.095  \\
       &   5   &  -0.042 &     0.198  &    0.060  \\
Local ridgelets ($B = 16$)   &  1   &    0.013    &    0.029     &   0.124  \\
       &  2   &   0.008    &    0.048     &   0.067 \\
Local ridgelets ($B = 32$)   &   1  &      0.062  &  0.030 &     0.060 \\
       &   2   &     0.058  &  0.044 &     0.114  \\
        &  3   &     0.040  &  0.106 &    0.0394 \\
Curvelets ($B = 16$)   &   1,1  &     0.117   &   0.016     &    10.120  \\
      &   1,2   &   0.020  &   0.029 &  0.010 \\
      &   1,3   &   0.017   &   0.094  &  0.077 \\
      &   2,1   &   0.064   &   0.023  &   0.207 \\
      &   2,2    &  0.065    &   0.038 &  0.019 \\
      &   2,3    &  0.058    &   0.056 &  0.029 \\
      &   2,4   &   0.05   &     0.175 &   0.072 \\
      &   3,1    &  0.060  &   0.034 &   0.084  \\
      &   3,2     & 0.060   &   0.045  &  0.082  \\
      &   3,3     & 0.058    &  0.073  &  0.016 \\
      &   3,4    &  0.028   &    0.196 &    0.251 \\
      &   4,1    &   0.257   &   0.073  &  0.009 \\
      &   4,2    &   0.261  &  0.083  &  0.038 \\
      &   4,3    &   0.276   &   0.109   &  0.080  \\
      &   4,4    &   0.260 &   0.250 &  0.051  \\
      &   5,1    &   0.262 &   0.146 &  0.016  \\
      &   5,2    &   0.272 &  0.149  &  0.017 \\
      &   5,3    &   0.274 &  0.187  &  0.041 \\  \hline
\end{tabular}
\caption{Table of mean excess kurtosis and standard deviation for 
$D^{(1)}=$CMB+KSZ. }
% \vspace{0.5cm}
\label{taball_kur_cmb_sz}
\end{center}
\end{table}
}

{\tiny
\begin{table}[htb]
\baselineskip=0.4cm
\begin{center}
\begin{tabular}{lcccc} \hline \hline
Multi-scale Method   & Scale  & ${\bar K_a}$ &$\sigma_K$ &  ${\cal K}$   \\ \hline \hline
Bi-orthogonal Wavelet  &  1,1 &       18.369 &        0.292 &   756.618 \\ 
 &        1,2 &       37.797 &        0.763 &     1550.101 \\ 
 &        1,3 &       37.960 &        0.139 &     1813.615 \\ 
 &        2,1 &        1.662 &        0.179 &       36.929 \\ 
 &        2,2 &        1.823 &        0.204 &       35.329 \\ 
 &        2,3 &       10.051 &        0.380 &      261.331 \\ 
 &        3,1 &        0.082 &        0.085 &        0.820 \\ 
 &        3,2 &        0.098 &        0.105 &        1.052 \\ 
 &        3,3 &        0.257 &        0.136 &        2.683 \\ 
 &       4,1 &       -0.023 &        0.181 &        0.064 \\ 
 &       4,2 &        0.034 &        0.194 &        0.476 \\ 
 &       4,3 &       -0.001 &        0.165 &        0.032 \\ 
 &       4,1 &       -0.060 &        0.354 &        0.097 \\ 
 &       4,2 &       -0.111 &        0.463 &        0.059 \\ 
 &       4,3 &       -0.037 &        0.479 &        0.086 \\ 
\`a trous WT &        1 &        5.744 &        0.153 &      424.159 \\ 
 &        2 &        0.250 &        0.049 &       11.003 \\ 
 &        3 &        0.030 &        0.055 &        0.688 \\ 
 &        4 &       -0.004 &        0.097 &        0.233 \\ 
 &        5 &       -0.032 &        0.191 &        0.271 \\ 
Local ridgelets ($B = 16$) &        1 &        0.174 &        0.034 &        5.682 \\ 
 &        2 &        0.055 &        0.049 &        0.773 \\ 
Local ridgelets ($B = 32$) &        1 &        0.153 &        0.031 &        2.840 \\ 
 &        2 &        0.082 &        0.047 &        0.413 \\ 
 &        3 &        0.066 &        0.113 &        0.024 \\ 
Curvelets ($B = 16$)  &        1,1 &        2.222 &        0.055 &      198.625 \\ 
 &        1,2 &        0.157 &        0.044 &        5.248 \\ 
 &        1,3 &        0.059 &        0.102 &        0.725 \\ 
 &        2,1 &        0.206 &        0.028 &        7.591 \\ 
 &        2,2 &        0.100 &        0.042 &        1.108 \\ 
 &        2,3 &        0.072 &        0.062 &        0.265 \\ 
 &        2,4 &        0.053 &        0.172 &        0.197 \\ 
 &        3,1 &        0.080 &        0.033 &        0.820 \\ 
 &        3,2 &        0.075 &        0.048 &        0.565 \\ 
 &       3,3 &        0.067 &        0.082 &        0.127 \\ 
 &       3,4 &        0.007 &        0.180 &        0.123 \\ 
 &       4,1 &        0.258 &        0.067 &        0.115 \\ 
 &       4,2 &        0.261 &        0.073 &        0.080 \\ 
 &       4,3 &        0.278 &        0.108 &        0.153 \\ 
 &       4,4 &        0.278 &        0.269 &        0.092 \\ 
 &       5,1 &        0.271 &        0.132 &        0.052 \\ 
 &       5,2 &        0.285 &        0.142 &        0.034 \\ 
 &       5,3 &        0.290 &        0.186 &        0.023 \\ \hline
\end{tabular}
\caption{Table of mean excess kurtosis and standard deviation for 
$D^{(2)}=$CMB+CS. }
% \vspace{0.5cm}
\label{taball_kur_cmb_cs}
\end{center}
\end{table}
}

{\tiny
\begin{table}[htb]
\baselineskip=0.4cm
\begin{center}
\begin{tabular}{lccccccc} \hline \hline
Multi-scale Method        & Scale        & ${\bar K_a}$ &$\sigma_K$ &  ${\cal K}$   \\ 
\hline \hline
Bi-orthogonal Wavelet  &        1,1 &       12.264 &        0.194 &      491.426 \\ 
 &        1,2 &       26.225 &        0.532 &     1040.543 \\ 
 &        1,3 &       20.180 &        0.069 &      933.862 \\ 
 &        2,1 &        1.503 &        0.160 &       34.154 \\ 
 &        2,2 &        1.697 &        0.187 &       33.660 \\ 
 &        2,3 &        7.477 &        0.291 &      196.930 \\ 
 &        3,1 &        0.092 &        0.089 &        0.922 \\ 
 &        3,2 &        0.095 &        0.103 &        1.027 \\ 
 &        3,3 &        0.258 &        0.130 &        2.749 \\ 
 &       4,1 &       -0.025 &        0.182 &        0.073 \\ 
 &       4,2 &        0.024 &        0.195 &        0.425 \\ 
 &       4,3 &       -0.000 &        0.166 &        0.035 \\ 
 &       5,1 &       -0.072 &        0.341 &        0.072 \\ 
 &       5,2 &       -0.110 &        0.462 &        0.062 \\ 
 &       5,3 &       -0.041 &        0.470 &        0.076 \\ 
\`a trous WT  &        1 &        5.206 &        0.127 &      392.254 \\ 
 &        2 &        0.260 &        0.051 &       11.972 \\ 
 &        3 &        0.032 &        0.055 &        0.747 \\ 
 &        4 &       -0.004 &        0.098 &        0.233 \\ 
 &        5 &       -0.033 &        0.187 &        0.280 \\ 
Local ridgelets ($B = 16$)   &        1 &        0.171 &        0.032 &        5.882 \\ 
 &        2 &        0.056 &        0.051 &        0.791 \\ 
Local ridgelets ($B = 32$)  &        1 &        0.150 &        0.031 &        2.988 \\ 
 &        2 &        0.080 &        0.047 &        0.392 \\ 
 &        3 &        0.063 &        0.114 &        0.006 \\ 
Curvelets ($B = 16$) &        1,2 &        0.155 &        0.044 &        5.515 \\ 
 &        1,3 &        0.058 &        0.099 &        0.717 \\ 
 &        2,1 &        0.197 &        0.029 &        7.426 \\ 
 &        2,2 &        0.099 &        0.041 &        1.091 \\ 
 &        2,3 &        0.071 &        0.061 &        0.254 \\ 
 &        2,4 &        0.049 &        0.173 &        0.172 \\ 
 &        3,1 &        0.079 &        0.033 &        0.803 \\ 
 &        3,2 &        0.073 &        0.046 &        0.519 \\ 
 &       3,3 &        0.067 &        0.081 &        0.128 \\ 
 &       3,4 &        0.004 &        0.178 &        0.135 \\ 
 &       4,1 &        0.258 &        0.068 &        0.121 \\ 
 &       4,2 &        0.258 &        0.075 &        0.047 \\ 
 &       4,3 &        0.272 &        0.105 &        0.108 \\ 
 &       4,4 &        0.276 &        0.261 &        0.076 \\ 
 &       5,1 &        0.270 &        0.133 &        0.053 \\ 
 &       5,2 &        0.284 &        0.144 &        0.028 \\ 
 &       5,3 &        0.287 &        0.186 &        0.016 \\ \hline
\end{tabular}
\caption{Table of mean excess kurtosis and standard deviation for 
$D^{(3)}=$CMB+KSZ+CS. }
% \vspace{0.5cm}
\label{taball_kur_cmb_sz_cs}
\end{center}
\end{table}
}

{\small
\begin{table}[htb]
\baselineskip=0.4cm
\begin{center}
\begin{tabular}{lcccc} \hline \hline
Kurtosis product                  & band & CMB+CS & CMB+SZ  &  CMB+SZ+CS   \\ \hline \hline
${\cal K}_{AT} * {\cal K}_{CUR}$& 1- 1,1 &    84248.508 &      665.882 &    64990.101 \\ 
       & 1- 1,2 &     2225.881 &        0.702 &     2163.283 \\ 
       & 1- 1,3 &      307.422 &        5.079 &      281.125 \\ 
       & 2- 2,1 &     2185.421 &       12.246 &     1983.507 \\ 
       & 2- 2,2 &       57.740 &        0.013 &       66.024 \\ 
       & 2- 2,3 &        7.975 &        0.093 &        8.580 \\ 
       & 2- 2,4 &       83.517 &        0.251 &       88.906 \\ 
       & 3- 3,1 &      136.694 &        0.295 &      123.816 \\ 
       & 3- 3,2 &        3.612 &        0.000 &        4.121 \\ 
       & 3- 3,3 &        0.499 &        0.002 &        0.536 \\ 
       & 3- 3,4 &        5.224 &        0.006 &        5.550 \\ 
       & 4- 4,1 &       46.342 &        0.966 &       38.613 \\ 
       & 4- 4,2 &        1.224 &        0.001 &        1.285 \\ 
       & 4- 4,3 &        0.169 &        0.007 &        0.167 \\ 
       & 4- 4,4 &        1.771 &        0.020 &        1.731 \\ 
       & 5- 5,1 &       53.779 &        0.610 &       46.327 \\ 
       & 5- 5,2 &        1.421 &        0.001 &        1.542 \\ 
       & 5- 5,3 &        0.196 &        0.005 &        0.200 \\  \hline \hline
\end{tabular}
\caption{Kurtosis product.}
% \vspace{0.5cm}
\label{taball_kur_mult}
\end{center}
\end{table}
}

\end{document}